\documentclass[prl,twocolumn,showpacs,amsmath,amssymb]{revtex4}
\input epsf

\usepackage{graphicx}

\newcommand{\la}{\langle}
\newcommand{\ra}{\rangle}
\newcommand{\Tr}{{\rm Tr}}
\newcommand{\be}{\begin{equation}}
\newcommand{\ee}{\end{equation}}
\newcommand{\bea}{\begin{eqnarray}}
\newcommand{\eea}{\end{eqnarray}}
\newcommand{\nl}{\nonumber \\}

\newcommand{\OO}{{\cal O}}

\newcommand{\Ref}[1]{(\ref{#1})}

\def\gsim{\mbox{~{\raisebox{0.4ex}{$>$}}\hspace{-1.1em}
	{\raisebox{-0.6ex}{$\sim$}}~}}
\def\lsim{\mbox{~{\raisebox{0.4ex}{$<$}}\hspace{-1.1em}
	{\raisebox{-0.6ex}{$\sim$}}~}}
\def\sss{\scriptscriptstyle}
\def\ca{C_{\sss A}}
\def\ch{C_{\!\sss H}}

\def\dh{d_{\sss H}}
\def\nf{N_{\rm f}}
\def\nc{N_{\rm c}}
\def\mD{m_{\!\sss D}}
\def\gs{g_{\rm s}}
\def\alphas{\alpha_{\rm s}}
\def\Eq#1{{\rm Eq}.~(\ref{#1})}
\def\q{{\bf q}}

\def\p{{\bf p}}
\def\putbox#1#2{{\epsfxsize=#1\textwidth \epsfbox{#2}}}
\def\centerbox#1#2{\centerline{\epsfxsize=#1\textwidth \epsfbox{#2}}}

\begin{document}
\title{Heavy quark diffusion in perturbative QCD at next-to-leading order}

\author{Simon Caron-Huot and  Guy D.~Moore}

\affiliation{Physics Department, McGill University, 3600
rue University, Montr\'eal, QC H3A 2T8, Canada}%

\begin{abstract}
We compute the momentum diffusion coefficient
of a nonrelativistic heavy quark in a hot QCD plasma,
to next-to-leading order in the weak coupling expansion.
Corrections arise at $\OO(\gs)$; physically they represent
interference between overlapping scatterings,
as well as soft, electric scale ($p\sim gT$) gauge field
physics, which we treat using the hard thermal loop
(HTL) effective theory.
In 3-color, 3-flavor QCD, the momentum diffusion constant of a
fundamental representation heavy quark at NLO is
$\kappa = \frac{16\pi}{3} \alphas^2 T^3 ( \ln \frac{1}{\gs} +
0.07428
+ 1.8869 \gs)$.  The convergence of the weak coupling expansion is poor.
\end{abstract}


\maketitle

The experimental program at RHIC and the future heavy ion program at the
LHC are exploring the behavior of the QCD plasma at temperatures above
the ``deconfinement'' temperature of $\sim 170$ MeV.  So far the
evidence is for a medium which interacts more strongly and thermalizes
more quickly than expected.  For instance, experimental results on
elliptic flow are well explained by hydrodynamics
\cite{hydro_works} but only if the shear viscosity is much less than a
naive extrapolation of weak coupling calculations
\cite{AMY6,Kapusta}.  Similarly, heavy quarks display
substantial elliptic flow and a degraded
energy spectrum \cite{heavy_xpt}, implying stronger medium
interactions than extrapolated weak-coupling calculations can easily
accommodate \cite{heavy_theory}.

This raises the general question; how well can we trust weak coupling
calculations for dynamical quantities in hot QCD at couplings anywhere
close to those probed in experiments?  Naively the perturbative series
converges in (possibly non-integer) powers of the strong coupling,
$\alphas \equiv \frac{\gs^2}{4\pi} \sim 0.4$ for relevant temperatures.
But perturbative series
often show convergence which is much better or much worse than one would
guess from the value of the coupling.  In general determining how well a
perturbative expansion converges requires evaluating a few terms in the
expansion.  Unfortunately, {\em no} dynamical transport quantity in QCD
which involves large length or time scales (such as shear and bulk
viscosity, electric conductivity, photon production, and heavy quark
momentum diffusion and energy loss) is known beyond leading order.
Most of the leading-order calculations are recent and quite involved.

Here we present a next-to-leading order calculation of the
theoretically simplest of these quantities, the momentum diffusion
coefficient of a nonrelativistic heavy quark.  This coefficient
(partially) characterizes how quickly heavy quarks are thermalized and
swept up in the collective flow of the plasma.


A heavy quark, $M \gg T$, in or near equilibrium has a typical momentum
$p\sim \sqrt{MT} \gg T$
large compared to the plasma scale and it therefore takes a
parametrically long time for the momentum to change appreciably.  This
means that momentum changes accumulate from many uncorrelated ``kicks,''
so on long time scales $p$ will evolve via Langevin dynamics,
\be 
\frac{d p_i}{dt} = -\eta_D\, p_i + \xi_i(t) \,,\quad 
\la \xi_i(t) \xi_j(t')\ra = \kappa\,\delta_{ij} \delta(t-t') \,.
\label{Langevin}
\ee
The relaxation rate $\eta_D$ and the momentum diffusion constant
$\kappa$ are
related by a fluctuation-dissipation relation,
$\eta_D = \frac{\kappa}{2MT}$,
which follows on general thermodynamical grounds.
Thus the dynamics of the nonrelativistic heavy quark
is completely set by the single parameter $\kappa$.
This parameter can be obtained by computing
the mean squared momentum transfer per unit time
in the  underlying microscopic theory.
In gauge theory, this mean squared momentum transfer equals
the time integrated correlator of two electric field operators
connected by fundamental Wilson lines \cite{CasalderreySolana}:
\be 
\kappa \equiv \frac{g^2}{3\dh}
  \!\!\int_{-\infty}^\infty \!\!\!\!\!\!dT\,
\Tr_H \la W(T;0)^\dagger \,E_i^{a}(T)t^a_H 
\,W(T;0) \,E_i^{b}(0)t^b_H \ra,
\label{defkappa1}
\ee
where $W(T;0)$ denotes a fundamental Wilson line
running from $t=0$ to $T$ along the static trajectory
of the heavy quark and $\dh=3$ is the dimension of the heavy quark's
representation.

Intuitively, \Eq{defkappa1} is exactly the force-force correlator of
\Eq{Langevin}, with the forces given by electric fields and the Wilson
line representing the gauge rotation of the heavy quark due to
propagation, which ensures
gauge invariance.  Because of operator ordering
issues, the Wilson lines shown are not equivalent to connecting the $E$
fields with an adjoint Wilson line, and in fact such Wilson lines are
even required in QED (diffusion of ions in a QED plasma depends on
the ionic charge $Z$ in a more complicated way than $Z^2$ only because
of these Wilson lines, which account for the reaction of the plasma to
the presence of the charge).


We start by showing how this formula reproduces the well known
\cite{BraatenThoma} leading order momentum diffusion coefficient.
At this order, \Ref{defkappa1} simplifies
to a zero-frequency Wightman correlator of two $A^0$ fields
(the $A^i$ fields do not contribute to the electric
field operators at zero-frequency in covariant and Coulomb gauges%
\footnote{%
    In some gauges (such as temporal gauge), the gauge field correlation
    function is divergent or pathological in the small frequency limit.
    It is only in such gauges that the $A^i$ fields in the electric
    fields make nontrivial contributions.})
\be 
\Ref{defkappa1} \Rightarrow
\frac{\ch g^2}{3} \int \frac{d^3p}{(2\pi)^3} p^2\,G^{>\,00}(\omega=0,p)
\label{locorrelator},
\ee

\begin{figure}[t]
\centerbox{0.44}{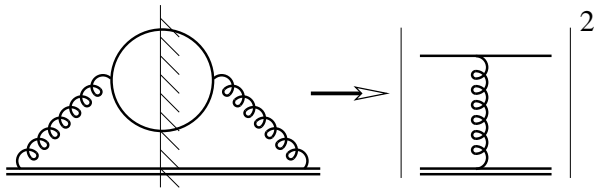}
\caption[leading-order diagram]
{\label{fig:relation} Leading-order contribution to heavy quark
  diffusion and its correspondence to scattering processes.  On the left
  the double line represents the Wilson line; on the right it is the
  heavy quark external states.}
\end{figure}

\noindent
where $\ch=\frac{4}{3}$ is the Casimir of the heavy quark's
representation.  This Wightman correlator can be evaluated in terms of
the squared matrix elements of t-channel scattering processes
involving the heavy quark, as illustrated in Fig.~\ref{fig:relation}.
These are the only processes which contribute in our case,
Compton-like processes being suppressed in the low velocity limit.
The result reduces to \cite{mooreteaney}
\bea \kappa^{\sss LO}
&\equiv& \frac{g^4\ch}{12\pi^3} \int_0^\infty q^2dq \int_{0}^{2q}
\frac{p^3\,dp}{(p^2+\mD^2)^2}
\nl &&
\times \left\{\begin{array}{l} 
N_f\,n_F(q)(1{-}n_F(q))\left( 2-\frac{p^2}{2q^2} \right)   \\
+N_c\,n_B(q)(1{+}n_B(q))  
        \left(2 -\frac{p^2}{q^2}+\frac{p^4}{4q^4} \right)\,.
\end{array} \right.
\label{defkappalo}
\eea
Here $p$ is the transferred momentum and $q$ is the energy of the light
scattering target.  Since the heavy quark is at rest, the initial and
final light-particle energies are equal and $p$ is purely spatial, which
is why the medium modification of the exchanged gluon propagator is
purely Debye screening with a Debye mass
$\mD^2=g^2T^2(N_c+N_f/2)/3$.  The inclusion of these HTL corrections
is essential for obtaining the complete leading order result,
otherwise $\kappa$ would be infrared divergent in the region
of soft momentum transfer $p$.
Formally taking $\mD \ll T$, the integral is dominated by $q \sim T$ and
$p$ in the parametric range $\mD \lsim p \lsim T$.
The strict leading-order evaluation of \Eq{defkappalo} yields
\be
\kappa \simeq
\frac{\ch g^4 T^3}{18\pi} \left[ \nc \left( \ln\frac{2T}{\mD} {+} \xi
  \right) + \frac{\nf}{2} \left( \ln\frac{4T}{\mD} {+} \xi \right)
  \right] \,,
\label{strictLO}
\ee
with $\xi = \frac{1}{2}-\gamma_{\sss E} + \frac{\zeta'(2)}{\zeta(2)}
\simeq -0.64718$.


When the exchange momentum $p$ is hard, $p \gsim T$, then
higher loop corrections to the propagators and vertices in
Fig.~\ref{fig:relation} represent $\OO(g^2)$ corrections.
However, the expression \Ref{defkappalo} for $\kappa$ receives
an $\OO(g)$ contribution from scatterings against soft gluons, $q \sim
\mD$.  Both the dispersion relations and the interactions of such gluons
are modified at the $\OO(1)$ level; at leading order these modifications
are described by hard thermal loops.  Therefore there will be $\OO(g)$
corrections to the above calculation.  But this is not the only source
of $\OO(g)$ next-to-leading order (NLO) corrections.

Another source is associated
with overlapping scattering events: the total scattering rate
for a hard particle is $\sim g^2T$, and is dominated by
t-Channel Coulombic scatterings involving soft momentum transfers.
These soft scatterings have a duration of order $\sim 1/\mD \sim 1/gT$
and therefore there is an $\OO(g)$ probability that two
such scattering events overlap with each other.
This is relevant in QCD (though not in QED, see below) because each
scattering color-rotates the participants.

\begin{figure}[tbh]
\hfill
\putbox{0.17}{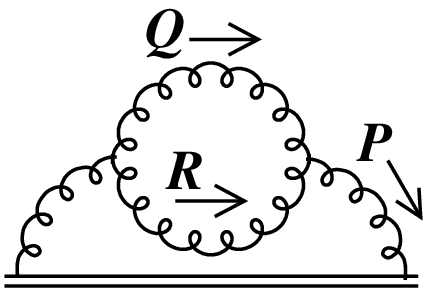}
\hfill \hfill
\putbox{0.13}{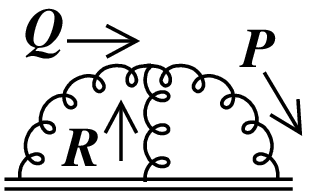}
\hfill $\phantom{.}$
\\
\hfill $(A)$ \hfill \hfill $(B)$ \hfill $\phantom{.}$
\\
\hfill
\putbox{0.13}{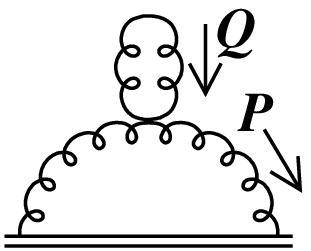}
\hfill \hfill
\putbox{0.13}{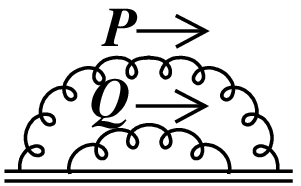}
\hfill $\phantom{.}$
\\
\hfill $(C)$ \hfill \hfill $(D)$ \hfill $\phantom{.}$
\\
\caption[NLO diagrams]
{\label{fig:diagrams} Diagrams required at NLO.  The double line is the
  Wilson line; otherwise all propagators are soft and HTL resummed and
  all vertices include the HTL vertex.  All lines attached to the Wilson
  line are longitudinal.}
\end{figure}

We need a systematic way of evaluating these NLO effects.  This is
provided by a loopwise expansion for \Eq{defkappa1}.  The diagrams
needed at NLO are shown in Fig.~\ref{fig:diagrams}.  The diagrammatic
series is convergent in powers of $g$ provided one incorporates HTL
corrections in propagators and vertices wherever momenta are soft
\cite{BraatenPisarski}, unless a diagram is sensitive to
the magnetic scale $\sim g^2 T$, which would be signaled by an infrared
divergence in the evaluation of a Feynman diagram.  This does not occur
in the current calculation; the diagrams shown in
Fig.~\ref{fig:diagrams} are all IR and UV convergent, after the
leading-order contribution is subtracted off from the transverse,
pole-pole contribution of diagram $(A)$.  Since the momenta are
soft, the ordering issues for the Wilson lines are subdominant and we
may replace the two Wilson lines in \Eq{defkappa1} with an adjoint
Wilson line; all diagrams involve the group theoretic combination $\ch
\ca$ and we may represent the NLO correction as the coefficient $C$
defined by
\be
\kappa \!=\! \frac{\ch g^4 T^3}{18\pi} \!
      \left( \!\left[\nc{+}\frac{\nf}{2}\right] \!\!
\left[\ln\frac{2T}{\mD}{+}\xi\right] {+} \frac{\nf \ln 2}{2}
{+} \frac{\nc \mD}{T} C \!\right)
\label{def_kappa}
\ee
with $\OO(g^2)$ corrections.
There is no $\OO(g)$ NLO correction in QED, where the (bare and HTL)
vertices involved in diagrams $(A),(B),(C)$ do not exist and the Wilson
line in $(D)$ is trivial.

Since \Eq{defkappa1} involves unequal time correlators we have found it
most convenient to evaluate it in the real-time (Schwinger-Keldysh)
formalism.  This required an extension of the HTL formalism to the
closed time path in the Schwinger-Keldysh ($r,a$) basis
\cite{huot}, which is convenient for treating soft physics because
Bose-Einstein factors only arise in one propagator.
We work in strict Coulomb gauge.  The measurable,
\Eq{defkappa1}, is gauge invariant and the HTL expansion should respect
gauge invariance, so we expect the sum of diagrams systematically
evaluated in powers of $g$ to produce gauge invariant results, though
the results for individual diagrams probably are not.

The effect of diagram $(A)$ can be divided into the real and the
imaginary part of the self-energy correction.  The real part is the
simplest to compute; it represents a correction to
the Debye mass which can actually be evaluated within the 3-D
dimensionally reduced theory \cite{dimred}.  4D Coulomb gauge corresponds
to 3D Landau gauge; in this gauge the self-energy receives a
nonzero, momentum-dependent contribution when one propagator in the
self-energy is transverse and the other is longitudinal ($A_0$ in the 3D
theory).  The correction is
\bea
\frac{1}{(p^2+\mD^2)^2} & \rightarrow &
   \frac{1}{(p^2 + \mD^2 + \delta \mD^2)^2} 
          \;\mbox{in Eq.~(\ref{defkappalo}),} \\
\delta \mD^2 & = & -4\ca g^2 T \int \frac{d^3 q}{(2\pi)^3}
\frac{p^2 - (\q\cdot\p)^2/q^2}{q^2 ((\p{-}\q)^2{+}\mD^2)}\,. \nonumber
\eea
The contribution to $C$ is found by expanding
$(p^2+\mD^2+\delta \mD^2)^{-2} - (p^2+\mD^2)^{-2}) \simeq -2 \delta \mD^2
(p^2+\mD^2)^{-3}$ and finding the shift to \Eq{defkappalo}.
Straightforward integration gives
$
C_{{\rm re}\:(A)} = \frac{3}{2\pi}\left(1+\frac{\pi^2}{16}\right)
\simeq 0.77199 \,.
$

The next simplest contributions are from diagrams $(C)$ and $(D)$.
Physically, $(C)$ accounts for real and virtual corrections in which the
light scatterer undergoes an additional soft scattering or soft plasmon
emission/absorption.  Diagram $(D)$ is the same but for the heavy
quark.  
In QED there is a cancellation between vertex orderings but in QCD
one instead picks up a commutator of color operators.
The contributions of these two diagrams are
\bea
C_{(C)} & \!\!=\!\! & 
 6\pi^2\!\! \int\!\! \frac{d^3 p}{(2\pi)^3} \frac{p^2}{(1+p^2)^2}
   \int\! \frac{d\Omega_v}{4\pi} \int\! \frac{d^4 Q}{(2\pi)^4} 
   G^{\mu\nu}_{rr}(Q) v_\mu v_\nu \nl
   && \hspace{3mm}
\times \frac{\delta(v\cdot (P{-}Q)) + \delta(v\cdot (P{+}Q))
     -2\delta(v\cdot P)}{(v\cdot Q)^2}\,,
\\
C_{(D)} & \!\!=\!\! & 
 \frac{3}{2\pi^4} \int_0^\infty p^4 dp
   \int_0^\infty q^2 dq \int_0^\infty d\omega \nl && \qquad \times
   \frac{G^{00}_{rr}(\omega,p)-G^{00}_{rr}(0,p)}{\omega^2}
   G^{00}_{rr}(\omega,q) \,.
\eea
In writing these expressions we have scaled all momenta by $\mD$ and
scaled out all powers of $T$.
Here $v^\mu\equiv(1,{\bf v})$ and
$G^{\mu\nu}_{rr}$ is the ordering-averaged gauge field correlator,
related to the retarded correlator via%
\footnote{We use the $[{-}{+}{+}{+}]$ metric and
define the retarded Green function without a factor of $i$, so at the
free level for a scalar field it is $-i/(Q^2-i\epsilon {\rm sgn}(q^0))$.}
$G^{\mu\nu}_{rr}(\omega,p) = (2n_{\sss B}(\omega){+}1)
\,{\rm Re}\: G^{\mu\nu}_{\sss R}(\omega,p) \simeq \frac{2T}{\omega}\,{\rm
  Re}\: G^{\mu\nu}_{\sss R}(\omega,p)$.
These expressions can be simplified somewhat but must be evaluated by
numerical quadratures.  We find \cite{HuotMoore2}
$C_{(C)} = -0.132916(1)$ and $C_{(D)} = 0.067526(1)$.

The most involved calculation is for the imaginary contribution of the
self-energy loop in diagram $(A)$.  This bears some similarity to the
calculation of the gluon damping rate by Braaten and Pisarski
\cite{gluondamp}, but the ``external'' momentum
$P$ is now spacelike. 
Therefore the integrals encountered are
four rather than two dimensional (one must integrate over $p$ and
$\theta_{pq}$), and the kinematics allow for processes
involving two soft plasmons on their mass shells, as well as virtual
corrections to the tree process of Fig.~\ref{fig:relation}.
The contribution to $C$ can be written
\begin{widetext}
\bea
\lefteqn{ \hspace{-0.2in}
C_{{\rm im}\:(A)} = 6\pi \int \frac{d^3p}{(2\pi)^3}
 \frac{p^2}{(1+p^2)^2} \int \frac{d^4Q}{(2\pi)^4}
\left[
-G_R^{\mu\mu'}\!(Q)\, G_R^{\nu\nu'}\!(R)\, 
   M_{\mu\nu}(Q,R) \,M_{\mu'\nu'}(Q,R)
\right. } && \nl && \hspace{1.6cm} \left.
{}+ 2G_{rr}^{\nu\nu'}\!(R) V_{\mu'\nu'}
\left(G_R^{\mu\mu'}\!(Q)\,  M_{\mu\nu}(Q,P)\,
     - G_A^{\mu\mu'}\!(Q)\, M_{\mu\nu}(Q,P)^*\right)
+ \frac12
V_{\mu\nu}V_{\mu'\nu'} \,G_{rr}^{\mu\mu'}\!(Q) \, G_{rr}^{\nu\nu'}\!(R)
\right] \,,\;\;
\eea
\end{widetext}
where we have introduced
\bea \hspace{-0.6cm}
M^{\mu\nu}(Q,R) &\equiv& \int \frac{d\Omega_v}{4\pi}
\frac{v^\mu v^\nu}{(v\cdot Q-i\epsilon)(v\cdot R-i\epsilon)} \,,
\\ \hspace{-0.6cm}
V_{\mu\nu} &\equiv& 2q^0\eta_{\mu\nu} + (R+P)_\mu \delta_\nu^0
-(Q+P)_\nu \delta_\mu^0
\eea
to denote objects that enter the HTL and tree vertices.
The evaluation is lengthy \cite{HuotMoore2},
and rather remarkably, turns out to be separable into pole-pole,
pole-cut and cut-cut contributions, 
in analogy to what was found by Braaten and Pisarski.
Two subtractions are required.
First, as mentioned above, at large momenta the pole-pole
contribution when both gauge bosons are transverse
duplicates the tree process
of Fig.~\ref{fig:relation}; this must be subtracted.
Further, evaluating the integrals in
\Eq{locorrelator} for finite $\mD$ already incorporates an NLO
correction besides what is in \Eq{strictLO}.  We will take the leading
contribution to be the result including this NLO correction
(which corresponds to $C_{\Eq{locorrelator}}=\frac{21}{8\pi}$.)
After these subtractions we obtain (numerically)
$C_{{\rm im}\:(A)} = 0.9097(1)$.

Diagram $(B)$ involves the correlator of three $A_0$ fields connected by
an HTL 3-point function (the tree vertex vanishes), and accounts for
interference between scattering events occurring 
on the light scatterer's side and on the heavy quark's side.
One of the $A^0$ fields carries zero frequency, and the contributions
can be organized according to whether the zero frequency propagator
is cut or retarded,
\bea
\hspace{-.2in}C_{\rm re\:(B)} &\!\!=\!\!&
6\pi^2\!\!\int \!\!\frac{d^3p}{(2\pi)^3} \frac{p}{(1{+}p^2)^2} \!\int\!\!
\frac{d^3q}{(2\pi)^3} \frac{M^{00}(q,-r)}{(1{+}q^2)(1{+}r^2)}
\,, \\
\hspace{-.2in} C_{\rm im\:(B)} 
  &\!\!=\!\!& 12\pi \int \frac{d^3p}{(2\pi)^3}
\frac{p^2}{1+p^2} \int \frac{d^4Q}{(2\pi)^4}
G_{rr}^{00}(R) \nl && \times \frac{G_R^{00}(Q)M^{00}(Q,p)-
G_A^{00}(Q)M^{00}(Q,p)^*}{q^0}\,.
\eea
The contributions $C_{\rm re\:(B)}$ and $C_{\rm im\:(B)}$
are closely analogous to the real and imaginary parts
of diagram $(A)$, respectively.
We find \cite{HuotMoore2} $C_{\rm re\:(B)}=-0.04829(1)$ and
$C_{\rm im\:(B)}=-0.07338(1)$.

\medskip

\centerline{\bf Discussion}

\medskip

The heavy quark diffusion coefficient can be computed beyond leading
order in the weak coupling expansion.  The first corrections arise at
$O(\gs)$ and describe ``soft'' $p,\omega \sim \gs T$ physics including
interference between scatterings and plasma corrections to interaction
strengths.  The calculation requires the HTL effective theory.

\begin{figure}[tbh]
\centerbox{0.45}{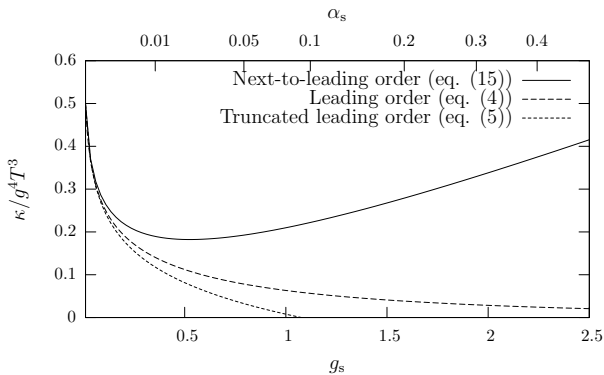}
\caption[Results]
{\label{fig:result}Comparison of leading and NLO results for $\nf=3$ QCD
  as a function of coupling.}
\end{figure}

The ratio of the NLO correction to the LO result is independent of
the representation of the heavy quark and is proportional to the
group's adjoint Casimir $\ca$ ($\ca=\nc$ in SU($\nc$) gauge theory, 0 in
QED).  Numerically, we find (see \Eq{def_kappa}) 
$C=1.4946+C_{\Eq{locorrelator}}$,
or, for 3 flavors,
\be
\kappa = \frac{16\pi}{3}\alphas^2 T^3 \left( \ln \frac{1}{\gs} +
0.07428 + 1.9026 \, \gs + \OO(\gs^2) \right) \,.
\label{eq:result}
\ee
The correction is positive, meaning faster equilibration of heavy
quarks.  As shown in Fig.~\ref{fig:result}, for realistic values of the
strong coupling the correction is large--a factor of 2 already at
$\alphas = 0.03$.  

Our result suggests that the convergence of the perturbative expansion
for dynamical quantities is poor.  Why?  About $\frac{1}{3}$ of the NLO
coefficient in \Eq{eq:result} (the part we called
$C_{\Eq{locorrelator}}$) is incorporated by integrating
\Eq{defkappalo} numerically rather than expanding it into
\Eq{strictLO}.  Another third, $C_{{\rm re}\:(A)}$, can be approximately
included into \Eq{defkappalo} by giving the real part of the self-energy
its full $p$ dependence rather than approximating it with
its small $p$ limit, $\mD^2$.  The remaining third
represents complicated and nontrivial many-body physics.

It would be interesting to make similar calculations for other transport
coefficients such as shear viscosity, and to extend the present
calculation to ${\cal N}{=}4$ SYM theory.

\smallskip

\centerline{\bf Acknowledgements}

\noindent
This work was supported in part by
the Natural Sciences and Engineering Research Council of Canada.

\end{document}